\documentclass[a4paper,twoside]{article}

\usepackage{epsfig}
\usepackage{subcaption}
\usepackage{calc}
\usepackage{amssymb}
\usepackage{amstext}
\usepackage{amsmath}
\usepackage{amsthm}
\usepackage{multicol}
\usepackage{pslatex}
\usepackage{apalike}
\usepackage[bottom]{footmisc}

\usepackage{hyperref}
\usepackage[utf8]{inputenc}

\usepackage{SCITEPRESS}     

\begin{document}

\title{Breathing Life into Models: The Next Generation of Enterprise Modeling}

\author{\authorname{Peter Fettke\sup{1}\sup{2}\orcidAuthor{0000-0002-0624-4431} and Wolfgang Reisig\sup{3}\orcidAuthor{0000-0002-7026-2810}}
\affiliation{\sup{1}German Research Center for Artificial Intelligence (DFKI), Saarbr\"ucken, Germany}
\affiliation{\sup{2}Saarland University, Saarbr\"ucken, Germany}
\affiliation{\sup{3}Humboldt-Universität zu Berlin, Berlin, Germany}
\email{peter.fettke@dfki.de, reisig@informatik.hu-berlin.de}
}

\keywords{systems composition, data modeling, behaviour modeling, process modeling, composition calculus, algebraic specification, systems mining}

\abstract{Edsger W. Dijkstra has frequently suggested building a “firewall” between the technology- and application- side of computer science. His justification: The methods to attack the computer scientists’ formal, mathematical “correctness problem” differ fundamentally from the methods to attack the applicants’ informal “pleasantness problem”. In this setting, a model is always confined to one side or the other of this wall. This keynote shows that a seamless transition between both sides can be achieved by a framework with \textit{architecture}, \textit{statics}, and \textit{dynamics} as the three pillars of modeling computer-integrated systems.  Selected examples justify this framework. It allows to “breath life” into (static) models, and it implies a new understanding of the “pleasantness” of computer-integrated systems, which is well-needed in the age of “digital first”.}

\onecolumn \maketitle \normalsize \setcounter{footnote}{0} \vfill

\section{\uppercase{Introduction: The two faces of computer science}}
\label{sec:introduction}

Computer-integrated systems exhibit two faces: the technological and the applied face. Edsger W. Dijkstra has frequently suggested to strictly separate both sides and to build a “firewall” between them \cite{dijkstra:firewallcacm}. His justification: The methods to attack the computer scientists' formal, mathematical “correctness problem” differ fundamentally from the methods to attack the applicants' informal “pleasantness problem”. In this setting, a model is always confined to one side or the other of this wall.

In contrast to Dijkstra, we understand modeling as an activity that should allow a seamless transition between formally and informally given or asserted facts of a computer-integrated system. Technology and applications must be interlocked by shared models, based on the same foundations. 

This keynote shows that our goal can be achieved by the \textsc{Heraklit} framework with \textit{architecture}, \textit{statics}, and \textit{dynamics} as the three pillars of modeling computer-integrated systems \cite{fettke2021handbook,fettke2021modelling}.  Selected examples justify this framework. It allows to “breathe life” into previous static, logic-based models, and it implies a new understanding of the “pleasantness” of computer-integrated systems. This is a key to understanding systems in the age of “digital first”.

This contribution consists of five sections. After this introduction, Section 2 explains what is needed for the foundations of computer-integrated systems from the application perspective. Section 3 presents our approach to digital modeling through a plain case study. Main characteristics of \textsc{Heraklit} and related work are discussed in Section 4 and 5, respectively. The paper concludes with Section 6.

\section{\uppercase{Towards foundations for the application side of computer science}}

In the last decades, computing-oriented disciplines have developed several strong ideas for the foundations of computer science. Typically, foundations of computer science stem from technical foundations such as abstractions of hardware, physical devices, sensors, actuators, etc. Another approach stems from theoretical computer science, starting with alphabets, formal languages, computable functions, etc.

But what are the foundations from the application-side of computer science? In this contribution, we argue that such foundations must cope with three fundamental phenomena:
\begin{itemize}
\item \textit{Architecture}: Computer-integrated systems in the real or an imagined world are not monolithic, amorphous, or unstructured, but can best be described and understood as the composition of different sub-systems. Hence, composing a system is inevitable for the proper understanding of a complex system. In one sentence: \textit{composition matters!}

\item \textit{Statics}: Computer-integrated systems process data. However, the world we live in consists not only of symbolic data objects but of many more relevant objects, e.g. invoices, customers, agreements, orders, and products. These objects need to be understood and represented adequately. Again, in one sentence: \textit{objects matter!}

\item \textit{Dynamics}: The world is not static, but continuously evolving. Events happen all over the world in time and space. A system’s progress in time induces an order of events; however, this order is irrelevant for a proper understanding of the event flow. On the contrary, it spoils the \textit{causal} order of events, which orders two events $a$ and $b$ by $a < b$ if and only if $a$ is a prerequisite for $b$. Of course, $a < b$ implies each potential clock to timestamp $a$ before $b$. But $a$ timestamped before $b$ only implies that $b$ is not a prerequisite for $a$. Again, in one sentence: \textit{causality matters!}
\end{itemize}

To sum up, composition, objects, and causality are important for understanding the application side of computer science. In other words, \textit{the world we are living in matters.} In our contribution, we show that \textsc{Heraklit}, an integrated modeling framework to formally describe the aspects mentioned before, is indeed feasible.

\section{\uppercase{A plain example}}

\begin{figure}[!h]
\centering
\includegraphics[scale=.18]{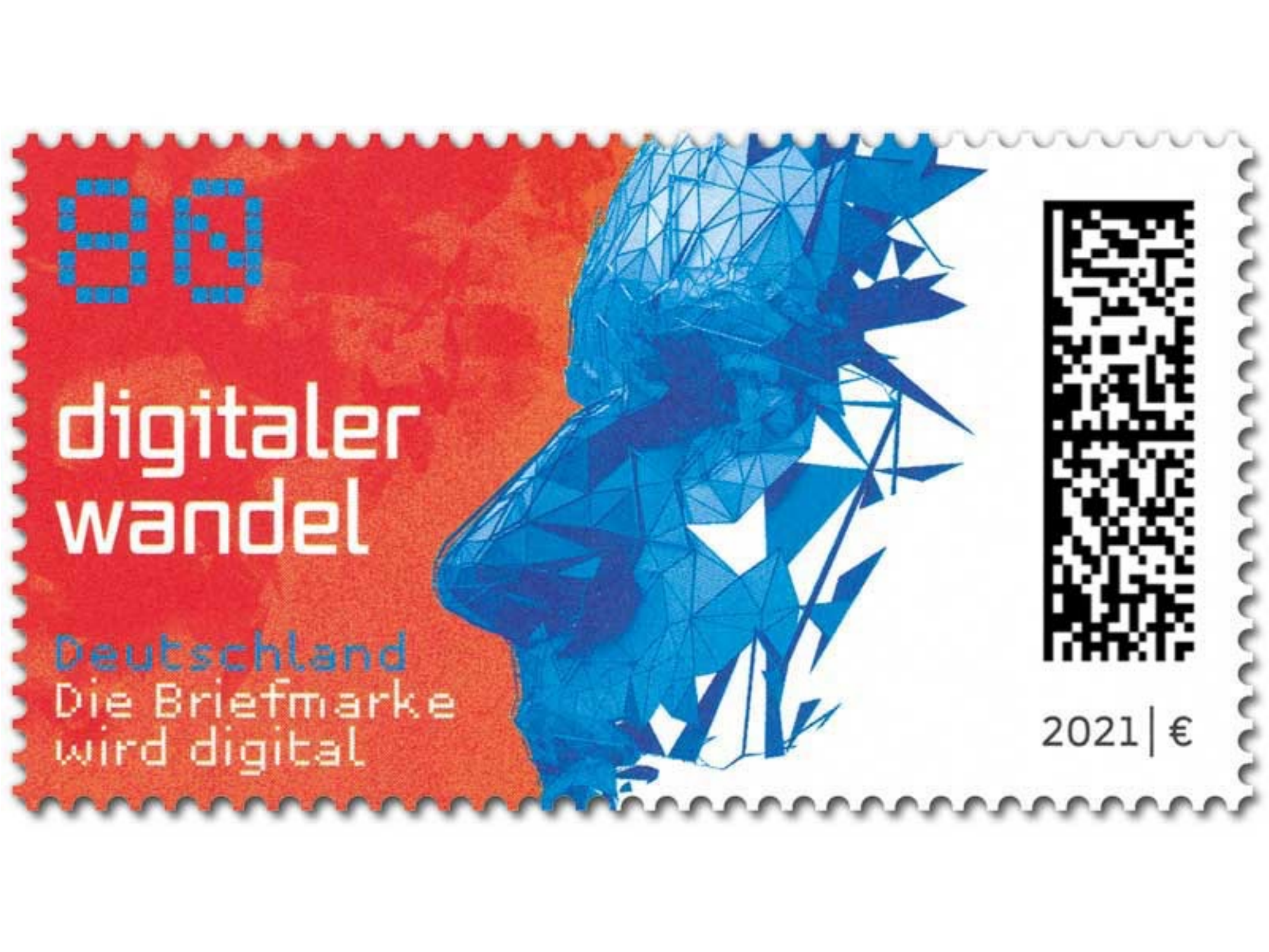}
\caption{Digital stamp with matrix code (right).}
\label{fig:digital_stamp}
\end{figure}

The following subsections illustrate the three \textsc{Heraklit}'s pillars by the plain example of a digital stamp, a stamp that is augmented with a matrix code. Fig.~\ref{fig:digital_stamp} depicts one exemplar of a digital stamp; each digital stamp carries a unique matrix code for individual identification.

\subsection{Architecture}

The basic concept of architecture is \textit{modules}. Inside a module, \textsc{Heraklit} typically uses Petri nets in various forms. In general, a module $M$ has two interfaces, the \textit{left interface} $^\ast M$, and the \textit{right interface} $M ^\ast$. An interface can contain Petri net \textit{places} as well as \textit{transitions}. Graphically, a module is represented as a rectangle, with the elements of the left interface on the left or top edge of the rectangle and the elements of the right interface on the right or bottom edge.

\begin{figure*}[!h]
\centering
\includegraphics[scale=.25]{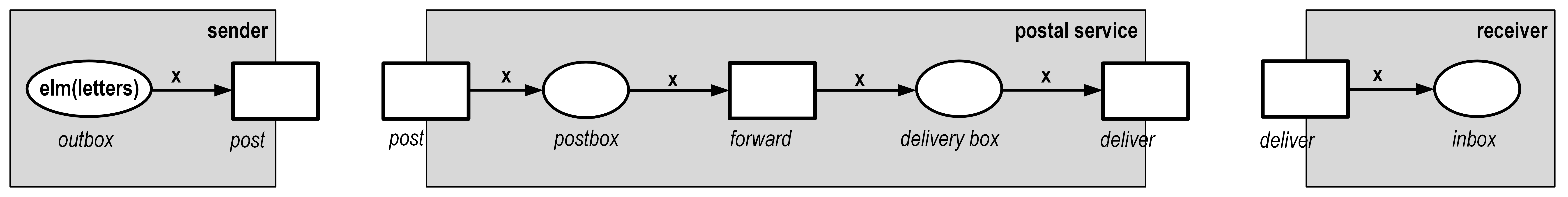}
\caption{Three modules: \textit{sender}, \textit{postal service}, and \textit{receiver}.}
\label{fig:three_modules}
\end{figure*}

Fig.~\ref{fig:three_modules} shows a typical model for the transportation of conventional letters. There are three modules: the \textit{sender}, the \textit{postal service}, and the \textit{receiver}. The left interface of the sender is empty, its right interface contains a Petri net transition. The label of this transition, \textit{post}, matches the labels of the transition of the left interface of the postal service; transitions with the same labels are merged later on when the two modules will be composed. Likewise, the right interface of the postal service and the left interface of the receiver contain equally labeled transitions.

The labels of the interface elements indicate the behavior and functionality of the modules: the sender posts each letter to the postal service. The postal service delivers each letter to the receiver. The receiver finds the letter in its inbox.

\begin{figure*}[!h]
\centering
\includegraphics[scale=.25]{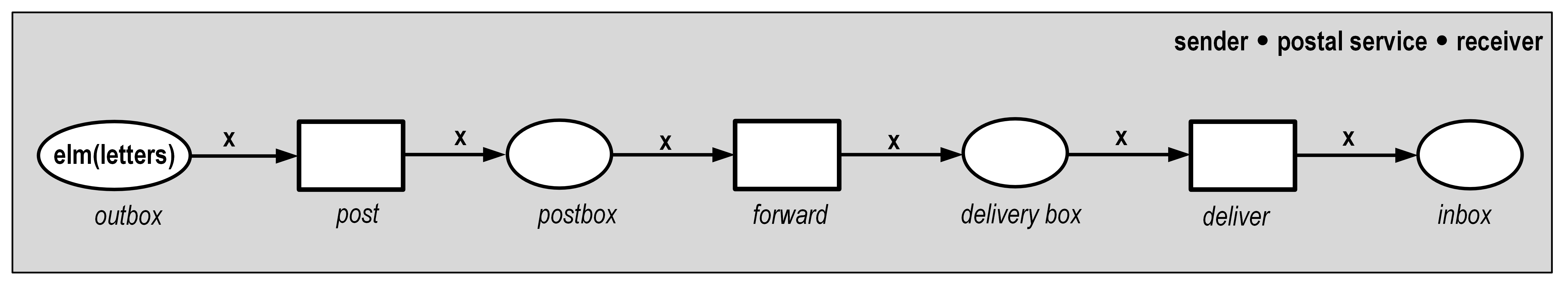}
\caption{The composition \textit{sender} $\bullet$ \textit{postal service} $\bullet$ \textit{receiver}.}
\label{fig:composition}
\end{figure*}

Fig.~\ref{fig:composition} shows the composition of the three modules, namely:
\begin{equation}
	 \textit{sender} \bullet \textit{postal service} \bullet \textit{receiver}
\end{equation}

with “$\bullet$” being the universal composition operator. This bracket free notation is possible since the composition operator is associative, i.e. for any three models $R, S, T$ holds \cite{reisig2019associative}:
\begin{equation}
    (R \bullet S) \bullet T = R \bullet (S \bullet T).
\end{equation}

\begin{figure*}[!h]
\centering
\includegraphics[scale=.25]{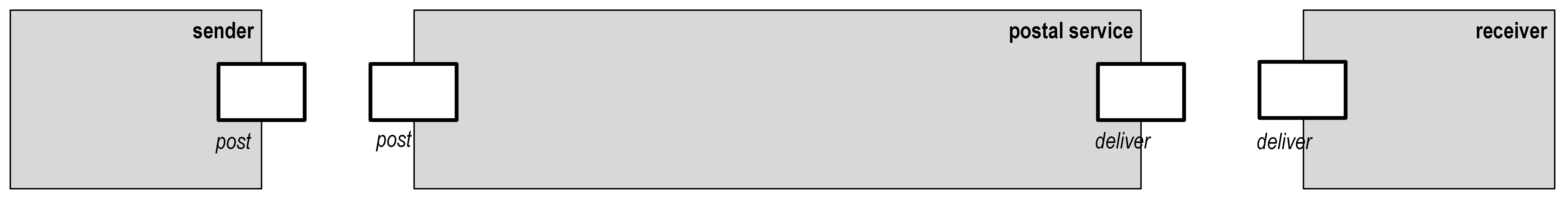}
\caption{The abstract form of the three modules \textit{sender}, \textit{postal service}, and \textit{receiver}.}
\label{fig:abstract_modules}
\end{figure*}

The greatest possible abstraction of the modules is depicted in Fig.~\ref{fig:abstract_modules}. Two activities of a postal service are modeled: post and deliver a letter; technically as Petri net transitions in $^\ast \textit{postal service}$ and $\textit{postal service} ^\ast$ with the labels \textit{post} and \textit{deliver}, respectively.

\begin{figure}[!h]
\centering
\includegraphics[scale=.35]{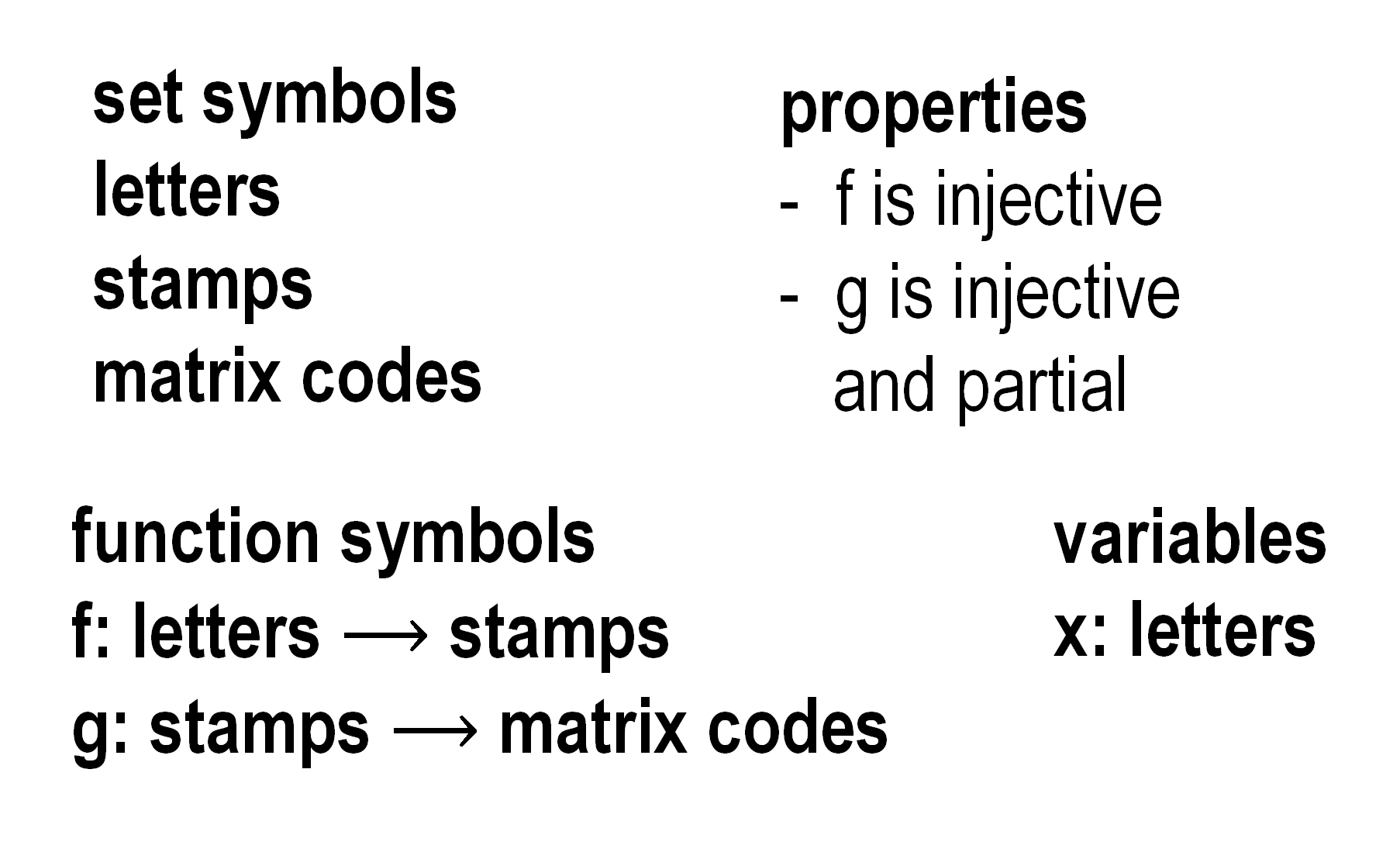}
\caption{Signature $\Sigma_0$.}
\label{fig:signature}
\end{figure}

\begin{figure}[h]
\centering
\includegraphics[scale=.35]{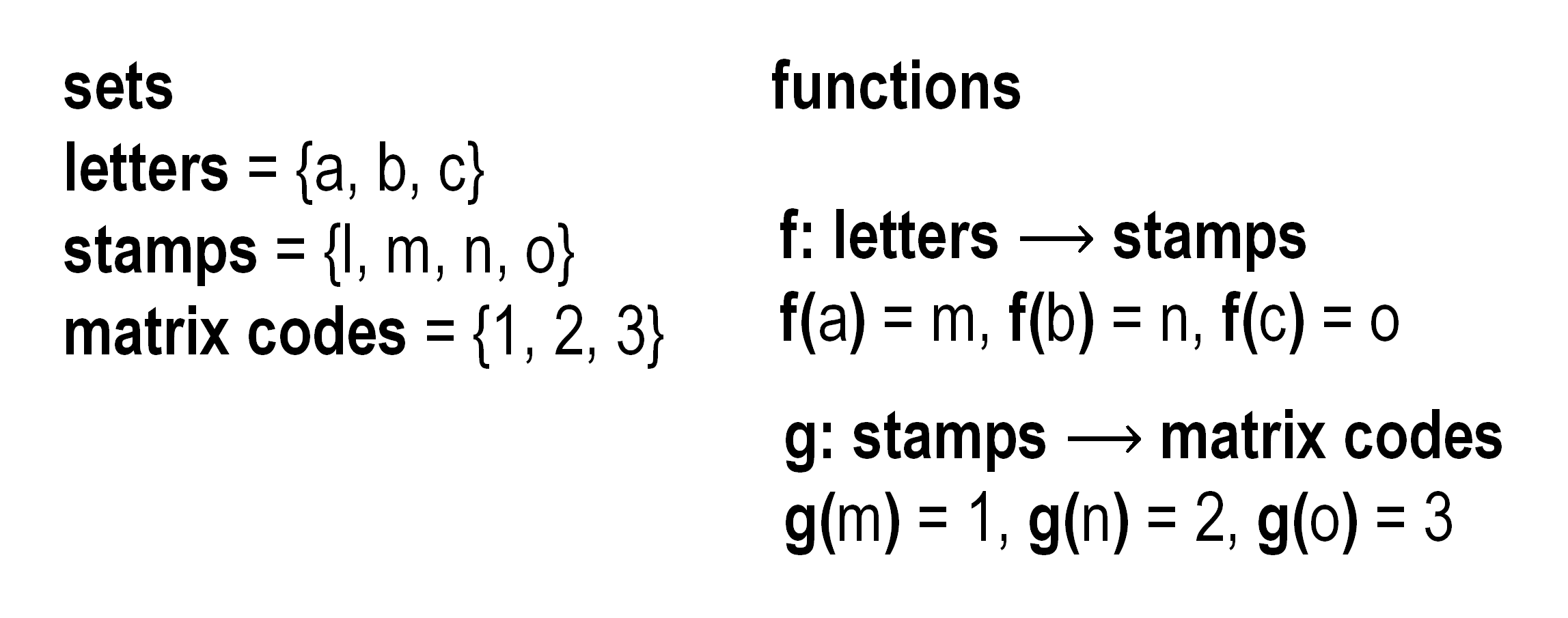}
\caption{$\Sigma$-structure $S_0$.}
\label{fig:structure}
\end{figure}

\subsection{Statics}

In each postal service, three sets of items play a central role: \textit{letters}, \textit{stamps}, and \textit{matrix codes}.  In a postal service, these quantities consists of different elements: each postal service has different sets of letters, stamps, and matrix codes.

Nevertheless, to formulate the processes for all postal services in the same way, we use a signature, $\Sigma_0$, shown in Fig.~\ref{fig:signature}, a technique to model a schema. This technique allows to characterize concrete objects from the real and imagined world as instantiations of such a schema. Here, we adopt notions such as structures, signatures, and instantiations of signatures, that are well-known from first-order logic \cite{suppes1957logic}. Technically, a signature is just a set of sorted symbols for \textit{sets}, \textit{constants}, and \textit{functions}. An instantiation interprets these symbols consistently. We extend signatures by requirements to exclude “unwanted” instantiations.

In the example, the signature contains the three symbols \textit{letters}, \textit{stamps}, and \textit{matrix codes} for the three sets described above, plus two symbols ($f$ and $g$) for functions. The symbol $f$ stands for a function that assigns to each letter its stamp; $g$ for a function that assigns to each stamp its matrix code. Both functions are injective. The function symbol $g$ stands for a function that is only partially defined since some stamps do not carry a matrix code.

The signature $\Sigma_0$ can be instantiated in many ways. Each concrete postal service is characterized by such an instantiation of the signature. In the case study, we discuss the instantiation $S_0$ of Fig.~\ref{fig:structure}. This instantiation consists of three letters $(a, b, c)$, four stamps $(l, m, n, o)$, three matrix codes $(1, 2, 3)$, and a particular assignment between letters, stamps, and matrix codes. For example, letter $a$ is assigned to stamp $m$; stamp $m$ is assigned to matrix code $1$.

\subsection{The behavior of the three modules at the schema level}

Signatures and their instantiations can naturally be trans\-ferred to define Petri net schemata. Such a sche\-ma can be instantiated in different ways; each instantiation results in a concrete Petri net.

We model the behavior of each of the three modules from Fig.~\ref{fig:three_modules}. Fig.~\ref{fig:composition} shows the behavior of each module at the schema level, i.e. based on the signature $\Sigma_0$, rather than a single instantiation. As a mathematical concept, a place represents a predicate that applies to a set of objects. This set can grow and shrink through the entry of transitions.

The sender module in Fig.~\ref{fig:three_modules} describes the letters to be sent. The question of an appropriate initial marking arises: for a given instantiation, i.e. a concrete postal service, the place \textit{outbox} initially contains all letters as tokens. In the example of the structure $S_0$ from Fig.~\ref{fig:structure}, these are the three tokens $a$, $b$, and $c$. On the schematic level, we only know the symbol \textit{letters}, for which each instantiation can freely choose an assignment with a set of letters. However, the intuitively obvious idea of labeling the place \textit{outbox} with the symbol \textit{“letters”} falls short: when instantiating the symbol \textit{letters} with a set, for example $\{a, b, c\}$, initially this set would emerge as a single token. However, we want the three elements of this set as individual tokens. This is noted as \textit{“elm(letters)”}.

Mathematically, the idea behind this is that a place represents a predicate that currently applies to the tokens on the place. The place \textit{outbox} with the inscription \textit{elm(letters)} thus stands for the logical expression:
\begin{equation}
\forall x \in \textit{letters} : \textit{outbox}(x).
\end{equation}

The instantiation $S_0$ thus generates in the initial marking the logical expression $\forall x \in \{a, b, c\} : \textit{outbox}(x)$. Thus, initially, the three tokens $a, b, c$ lie on the place \textit{outbox}.

In a given instantiation $S_0$, the transition \textit{post} can occur in the sender or postal service module as soon as a token is present on the place \textit{outbox}. The assignment of the variable $x$ can freely be chosen from the set with which the instantiation $S_0$ instantiates the set symbol \textit{letters}. In analogy, the postal service module forwards one letter from the postbox to the delivery box. Finally, the letter is delivered to the receiver.

\subsection{The instantiation $S_0$ and its behavior}

\begin{figure*}[!h]
\centering
\includegraphics[scale=.25]{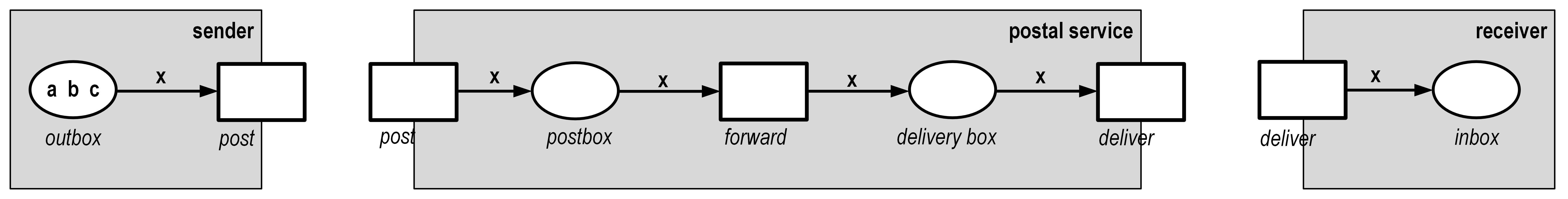}
\caption{Three instantiated modules.}
\label{fig:system_modules}
\end{figure*}

\begin{figure*}[!h]
\centering
\includegraphics[scale=.25]{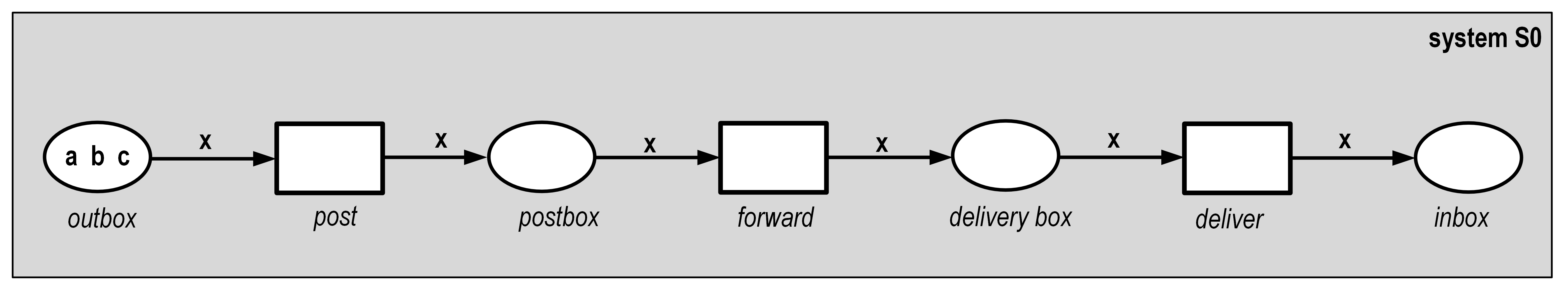}
\caption{System $S_0$.}
\label{fig:system_S0}
\end{figure*}

Fig.~\ref{fig:system_modules} shows three instantiated modules; Fig.~\ref{fig:system_S0} shows the module system $S_0$, which is created from the modules of Fig.~\ref{fig:three_modules} in two steps: first, the three modules from Fig.~\ref{fig:three_modules} are composed into a single module, and second, the signature $\Sigma_0$ of Fig.~\ref{fig:signature} is instantiated with the structure $S_0$ of Fig.~\ref{fig:structure}. In particular, the place \textit{outbox} now contains three tokens.

In the instantiation $S_0$, one can now document the processing of individual postal services. In the initial state, for example, the transition \textit{post} is activated. After that, the letter can be posted by the sender, forwarded to the delivery box, and be delivered to the receiver. 

\textsc{Heraklit} models a behavior as a distributed run; formally conceived as a module. This makes it easy to compose distributed runs. The current example also shows that distributed runs show details of behavior much more explicitly than is possible with sequential runs. In particular, relationships between objects and causality is explicitly modeled.

\begin{figure*}[!h]
\centering
\includegraphics[scale=.25]{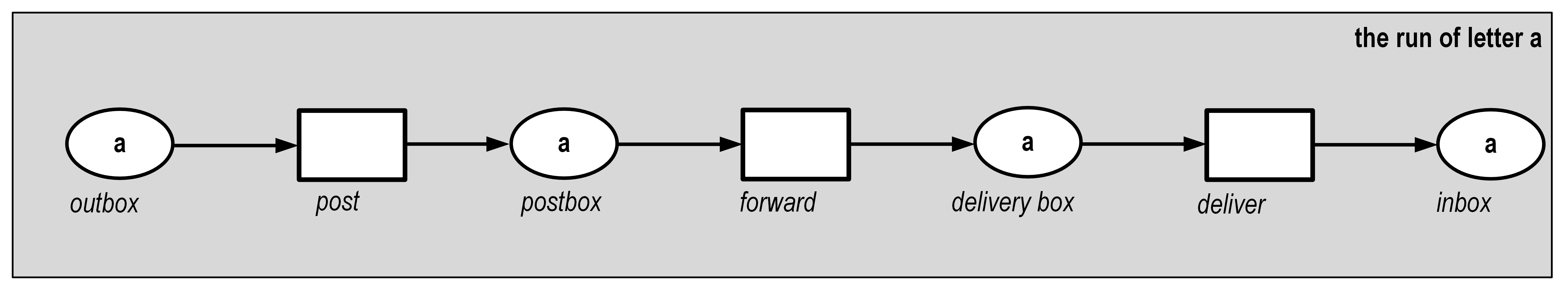}
\caption{The run of letter $a$.}
\label{fig:run_a}
\end{figure*}

\begin{figure*}[!h]
\centering
\includegraphics[scale=.25]{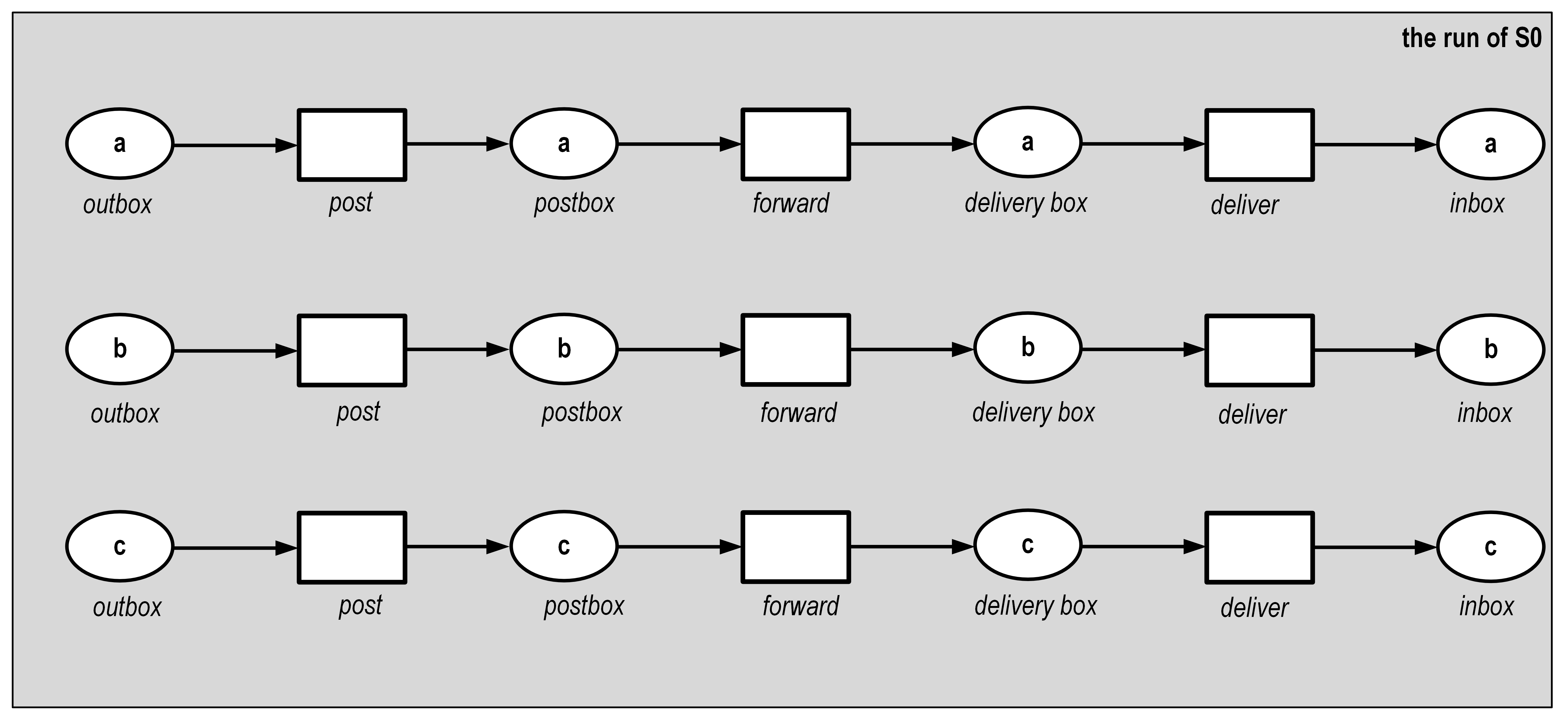}
\caption{The run of $S_0$.}
\label{fig:the_run}
\end{figure*}

Fig.~\ref{fig:run_a} shows what happens with the letter $a$: The letter moves from the sender's outbox to the postbox, then to the delivery box and finally to the receiver's inbox. Technically, this is the run of letter $a$ in the system $S_0$. Fig.~\ref{fig:system_S0} depicts two more letters, $b$ and $c$, yielding corresponding runs. Together, the three runs of $a$, $b$, and $c$ define the overall run of the system $S_0$, shown in Fig.~\ref{fig:the_run}.  In the run of each single letter, the events \textit{post}, \textit{forward} and \textit{deliver} occur strictly ordered. As the letters proceed independently of each other, the events of different letters occur independently, namely their occurrences are not ordered in the overall run of the system $S_0$.

Fig.~\ref{fig:the_run} exemplifies that events are not ordered along a time axis; instead, order represents causality. For example, the event \textit{forward} for letter $a$ occurs causally after the event \textit{post} for letter $a$, but is causally independent from all events of the letters $b$ and $c$. The run of Fig.~\ref{fig:the_run} is the only run of the system $S_0$ in Fig.~\ref{fig:system_S0}. System $S_0$ is deterministic, as there is never a decision to be taken; and $S_0$ is concurrent, as some events occur \textit{unordered}. One may nevertheless introduce global views into the run of Fig.~\ref{fig:the_run}. A global view is a maximal set of unordered places. Each of the three letters can be in exactly one of the four boxes. In technical terms, each such set corresponds to one of the 64 ($= 4 \times 4 \times 4$) reachable markings of the system in Fig.~\ref{fig:system_S0}.

\section{\uppercase{Main characteristics of \textsc{Heraklit}}}

\begin{figure}[!h]
\centering
\includegraphics[scale=.17]{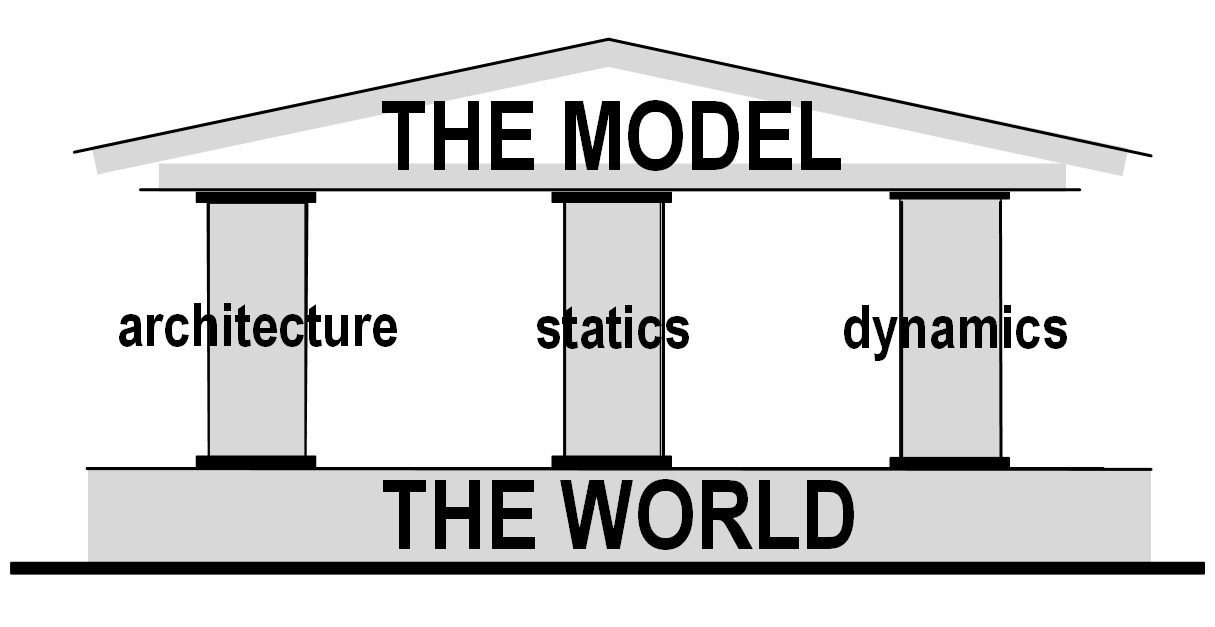}
\caption{The three pillars of \textsc{Heraklit}.}
\label{fig:temple}
\end{figure}

The main characteristics of the \textsc{Heraklit} framework were introduced by means of a plain example. From a conceptual point of view, the \textsc{Heraklit} framework is based on three pillars \cite{fettke2021handbook,fettke2021modelling}. Each pillar focuses on different but integrated aspects of modeling (Fig.~\ref{fig:temple}):
\begin{itemize}
\item Architecture: A \textsc{Heraklit} model consists of modules. Each module has an arbitrarily designed interior; in particular, the level of abstraction of the representation is freely selectable. The interface of a module contains labeled elements. The composition of modules is technically simple and always associative. The formal basis of this architectural principle is the composition calculus \cite{reisig2019associative}.

\item Statics: For dealing with data and data structures, \textsc{Heraklit} employs abstract data types and algebraic specifications, as they have turned out to be useful in computer science from the beginning and have long been used in specification languages such as $VDM$, $Z$, etc. \cite{sanella20212algebraic}. Symbolic representations (in terms of a signature) are used as in predicate logic. Each module has a signature, i.e. a set of symbols for sets, constants, and functions, from which, together with variables, terms are then formed. A signature, and thus its terms, can be instantiated in quite different ways.

\item Dynamics: A module for describing behavior contains a Petri net inside of it. Here, each place of the Petri net is a logical predicate, each arrow is labeled with a term of the module's signature, and each transition with a condition. For a given instantiation of the signature, the set of objects to which the predicate applies can grow or shrink due to the occurrence of an event. In detail, this is described by terms at the arrows of the Petri net. This allows behavior to be represented abstractly on a schematic level, but also concretely for a “meant” instantiation.
\end{itemize}

\section{\uppercase{Related work}}
In the last decades, a plethora of modeling frameworks emerged \cite{frank2014:modeling,sandkuhl2018modeling,vernadat2020enterprise,wand2002commentary}. In 1992 already, some colleagues coined the term “methodology jungle” for characterizing the variety and diversity of the research field \cite{hofstede1992jungle}. Nowadays, this field is still evolving and maturing. Relevant work can be approached from different angles, e.g. the problem addressed or the solution purported.

In the field of business informatics, typical frameworks used are \textit{Architecture for Integrated Information Systems} (ARIS) \cite{scheer1999ARIS}, \textit{Business Process Model and Notation} (BPMN) \cite{OMG2014BPMN}, \textit{Event-Driven Process Chains} (EPC) \cite{keller1992EPK}, MEMO \cite{frank2014memo}, the St. Gallen Approach to Business Engineering \cite{oesterle1995engineering,winter2001engineering}, and \textit{Unified Modeling Language} (UML) \cite{OMG2017UML}. Most of them offer graphical representations to describe behavior (e.g. BPMN, EPC), some have expressive means for data structures (e.g. ARIS, MEMO), and in some cases also for schemas (MEMO). Some known frameworks use more or less formally defined composition operations. None of the frameworks are fully formally defined but describe many aspects only intuitively. Thus, no other framework reaches the scope of expressive means of \textsc{Heraklit}. This characterization is also true for frameworks stemming from the information systems discipline, e.g. Bunge's ontological model \cite{wand1990ontological} or work system theory \cite{alter2013wst}.

From a computer science perspective, several modeling frameworks were proposed, e.g. \textit{Abstract State Machines} (ASM) \cite{gurevich2000ASM}, \textit{Aloy} \cite{jackson2019alloy}, \textit{High-level Petri Nets} \cite{genrich1981predicatenets,jensen2009colourednets}, \textit{event-B} \cite{abrial2005B}, \textit{FOCUS} \cite{broy1997refinement,broy2001specification}, \textit{Statecharts} \cite{harel1987statecharts}, \textit{TLA} \cite{lamport2002TLA}, and \textit{Z} \cite{spivey1992B}. The central ideas and concepts of \textsc{Heraklit} combine three well-proven, intuitively easy to understand, and mathematically sound concepts that have been used for the specification of systems in the past:
\begin{itemize}
\item Architecture: The composition calculus for structuring large systems. This calculus with its widely applicable associative composition operator is the most recent contribution to the foundations of \textsc{Heraklit}. The obvious idea, often discussed in literature, of modeling composition as fusion of equally labeled interface elements of modules is supplemented by the distinction of left and right interface elements, and composition $A \bullet B$ as fusion of right interface elements of $A$ with left interface elements of $B$. According to \cite{reisig2019associative}, this composition is associative (as opposed to the naive fusion of interface elements); it also has several other useful properties. In particular, the composition is compatible with refinement and with single (distributed) runs.

\item Statics: Abstract data types and algebraic specifications for the formulation of concrete and abstract data: since the 1970s such specifications have been utilized, built into specification languages, and often used for (domain-specific) modeling. \cite{sanella20212algebraic} presents systematically the theoretical foundations and some applications of algebraic specifications. 

\item Dynamics: Petri nets for formulating dynamic behavior: \textsc{Heraklit} uses the central ideas of Petri nets  \cite{petri1962diss,reisig2013understanding}. A step of a system, especially of a large system, has locally limited causes and effects \cite{petri1977non_sequential}. This allows processes to be described without having to use global states and globally effective steps. The connection with algebraic specifications is established by \cite{reisig1991algebraic}. 

\end{itemize}

These three theoretical principles harmonize together and generate further best-practice concepts that contribute to a methodical approach of modeling with \textsc{Heraklit} and which could only be touched upon in this keynote. On the downside, industrially mature tools for \textsc{Heraklit} are still under development.

\section{\uppercase{Conclusions}}
The fundamental focus of our approach becomes clear in comparison with the firewall propagated by Dijkskra: \textsc{Heraklit} follows the tradition of logic, integrating formal arguments as part of the scientific penetration of real-life facts. Logic has successfully been doing this for 2000 years. \textsc{Heraklit} extends formal logic by including dynamic aspects, and shapes a connection between informal and formal concepts. 

Like no other framework, \textsc{Heraklit} combines the three pillars of architecture, statics, and dynamics for modeling computer-integrated systems. Formulated in buzzword form on a technical level:
\begin{gather*}
    \textit{\textsc{Heraklit}} = \\ \textit{modules } + \textit{ relational structures }\ + \textit{ local steps},
\end{gather*}

and on the level of calculi:
\begin{gather*}
    \textit{\textsc{Heraklit}} = \\ \textit{composition calculus } + \textit{ logic }\ + \textit{ Petri nets.}
\end{gather*}

With \textsc{Heraklit}, both sides of Dijskrta's wall can be grounded on the same foundations. It is obvious that this approach has decisive advantages and will achieve tremendous gains: It is seamlessly possible to capture the main ideas of a natural and intuitive understanding of the world we live in, enrich this understanding with formal concepts, and use the description as a foundation for (forthcoming) supporting software.

The technical \textsc{Heraklit} concepts described can be used to model aspects of discrete systems in a way that no other modeling technique can:
\begin{itemize}
\item Modules of different hierarchical levels can be composed, e.g. a module may represent the behavior of a component at the finest, operational level. At the same time, its neighboring modules are given only by their names and their interface.

\item Real-world and abstract objects are coequal elements of formal reasoning. \textsc{Heraklit} follows predicate logic and conceives both, real-world and formal objects, as interpretations of terms of a signature (an alphabet with typed symbols). 

\item Behavioral models can be parameterized. Each interpretation of the symbolic labels of a schematic \textsc{Heraklit} module generates a separate system model. For example, a schematic module describes the principles of business processes of all branches of a commercial bank; each interpretation describes a concrete branch. 

\item The size of models grows linearly with the size of systems. This is especially true for modeling dynamic behavior. To achieve this, \textsc{Heraklit} works without globally effective constructs, in particular without global states.  The notion of single runs, as well as the composition operator for modules, are likewise locally confined.

\item \textsc{Heraklit} adopts verification techniques. Petri nets provide a variety of efficient computational verification techniques. The most important ones, in particular \textit{place-} and \textit{transition invariants}, also work at the symbolic (signature) level and -- more expressively -- for individual interpretations of a signature. Verification based on model checking is only possible for individual interpretations of a signature; efficient procedures exist for this. A large, still open research area is \textit{compositional verification}: properties of a composed system are derived from properties of the composed modules.

\item A single run can be distributed. In a run of a system, events often occur independently. For example, in a bank branch, employees interact with individual customers largely independently; they are only occasionally synchronized when accessing scarce resources, such as meeting rooms. Many modeling frameworks model this kind of independent events as “occurence in any order” . This leads to exponentially many supposedly different “sequential” processes. \textsc{Heraklit} explicitly models independence of events, not only in sequences but any other partial order. This yields a clear notion of non-determinism and a theorem according to which the runs of a composed system are derivable from the runs of its components.

\item \textsc{Heraklit} uses intuitively simple basic concepts: predicate logic is familiar to many users from other contexts anyway. Petri nets are also widely used in computer science and its application areas. The composition calculus is immediately conceivable from small examples. The graphical means of expression of Petri nets and the composition calculus complement each other harmoniously.
\end{itemize}

It is plausible that \textsc{Heraklit} will help to harmonize the plethora of different phenomena associated with computer-integrated systems. In other words, it will be possible to describe the essence of different application types, and system structures of applications, including real-time systems, reactive systems, data-intensive systems, cyber-physical systems, service-based systems, integrated business systems for enterprise resource planning (“ERP systems”), supply chain management (“SCM systems”), customer relationship (“CRM systems”, “ticket systems”), product lifecycle management (“PLM systems), and many more system types in the “digital world”.

In the future, more work is needed to develop the \textsc{Heraklit} framework into a universal modeling infrastructure:
\begin{enumerate}
\item More case studies and industrial applications are required, to identify special classes of models, shorthands, best practices, etc.

\item \textsc{Heraklit} specific analysis methods have to be developed, respecting and employing the local character of events and modules, to formulate and to prove decisive properties of models.

\item Powerful software tools have to be created, supporting the construction of models, their analysis, the integration of databases, etc.
\end{enumerate}

So far, \textsc{Heraklit} provides fundamental pillars for systematically developed, theory-based models, covering a broader spectrum of aspects than any other modeling infrastructure. It is now time to “breathe life” into models with \textsc{Heraklit}.

\bibliographystyle{apalike}
{\small
\bibliography{main}}

\begin{thebibliography}{}

\bibitem[Abrial, 2005]{abrial2005B}
Abrial, J.-R. (2005).
\newblock {\em The {B-Book} -- Assigning Programs to Meanings}.
\newblock Cambridge University.

\bibitem[Alter, 2013]{alter2013wst}
Alter, S. (2013).
\newblock Work system theory: Overview of core concepts, extensions, and
  challenges for the future.
\newblock {\em Journal of the Association for Information Systems}, 14(2).

\bibitem[Broy, 1997]{broy1997refinement}
Broy, M. (1997).
\newblock Compositional refinement of interactive systems.
\newblock {\em Journal of the ACM}, 44(6):850--891.

\bibitem[Broy and Stolen, 2001]{broy2001specification}
Broy, M. and Stolen, K. (2001).
\newblock {\em Specification and Development of Interactive Systems: Focus on
  Streams, Interfaces, and Refinement}.
\newblock Springer.

\bibitem[Dijkstra, 1989]{dijkstra:firewallcacm}
Dijkstra, E.~W. (1989).
\newblock Reply to comments.
\newblock {\em Commun. ACM}, 32(12):1414.

\bibitem[Fettke and Reisig, 2021a]{fettke2021handbook}
Fettke, P. and Reisig, W. (2021a).
\newblock Handbook of \textsc{Heraklit}.
\newblock \textsc{Heraklit} working paper, v1.1, September 20, 2021,
  \url{http://www.heraklit.org}.

\bibitem[Fettke and Reisig, 2021b]{fettke2021modelling}
Fettke, P. and Reisig, W. (2021b).
\newblock Modelling service-oriented systems and cloud services with
  \textsc{Heraklit}.
\newblock In Zirpins, C., Paraskakis, I., Andrikopoulos, V., Kratzke, N., Pahl,
  C., El~Ioini, N., Andreou, A.~S., Feuerlicht, G., Lamersdorf, W., Ortiz, G.,
  Van~den Heuvel, W.-J., Soldani, J., Villari, M., Casale, G., and Plebani, P.,
  editors, {\em Advances in Service-Oriented and Cloud Computing}, pages
  77--89, Cham. Springer International Publishing.

\bibitem[Frank, 2014]{frank2014memo}
Frank, U. (2014).
\newblock Multi-perspective enterprise modeling: foundational concepts,
  prospects and future research challenges.
\newblock {\em Journal of the Association for Information Systems},
  13(2):941--962.

\bibitem[Frank et~al., 2014]{frank2014:modeling}
Frank, U., Strecker, S., Fettke, P., vom Brocke, J., Becker, J., and Sinz,
  E.~J. (2014).
\newblock The research field ''modeling business information systems`` --
  current challenges and elements of a future research agenda.
\newblock {\em Business \& Information Systems Engineering}, 6(1):39--43.

\bibitem[Genrich and Lautenbach, 1981]{genrich1981predicatenets}
Genrich, H.~J. and Lautenbach, K. (1981).
\newblock System modelling with high-level petri nets.
\newblock {\em Theoretical Computer Science}, 13:109--135.

\bibitem[Gurevich, 2000]{gurevich2000ASM}
Gurevich, Y. (2000).
\newblock Sequential abstract-state machines capture sequential algorithms.
\newblock {\em ACM Transactions on Computational Logic}, 1:77--111.

\bibitem[Harel, 1987]{harel1987statecharts}
Harel, D. (1987).
\newblock Statecharts: A visual formalism for complex systems.
\newblock {\em Science of Computer Programming}, 8(3):231--274.

\bibitem[Jackson, 1987]{jackson2019alloy}
Jackson, D. (1987).
\newblock Alloy: A language and tool for exploring software designs.
\newblock {\em Communications of the ACM}, 62(9):66--76.

\bibitem[Jensen and Kristensen, 2009]{jensen2009colourednets}
Jensen, K. and Kristensen, L.~M. (2009).
\newblock {\em Coloured Petri Nets: Modelling and Validation of Concurrent
  Systems}.
\newblock Springer.

\bibitem[Keller et~al., 1992]{keller1992EPK}
Keller, G., Nüttgens, M., and Scheer, A.-W. (1992).
\newblock {S}emantische {P}rozeßmodellierung auf der {G}rundlage
  {E}reignisgesteuerter {P}roze{\ss}ketten ({EPK}).
\newblock Technical Report~89, Veröffentlichungen des Instituts für
  Wirtschaftsinformatik (IWi) an der Universität des Saarlandes.

\bibitem[Lamport, 2002]{lamport2002TLA}
Lamport, L. (2002).
\newblock {\em Specifying Systems: The TLA+ Language and Tools for Hardware and
  Software Engineers}.
\newblock Adison-Wesley.

\bibitem[{Object Management Group}, 2014]{OMG2014BPMN}
{Object Management Group} (2014).
\newblock Business process model and notation (bpmn): Version 2.0.2.
\newblock Technical Report formal/2013-12-09, Object Management Group.

\bibitem[{Object Management Group}, 2017]{OMG2017UML}
{Object Management Group} (2017).
\newblock Omg unified modeling language (omg uml): Version 2.5.1.
\newblock Technical Report formal/2017-12-05, Object Management Group.

\bibitem[Petri, 1962]{petri1962diss}
Petri, C.~A. (1962).
\newblock {\em Kommunikation mit Automaten}.
\newblock PhD thesis, Institut f\"ur instrumentelle Mathematik der
  Universit\"at Bonn.

\bibitem[Petri, 1977]{petri1977non_sequential}
Petri, C.~A. (1977).
\newblock Non-sequential processes.
\newblock Technical Report ISF-77-5, Gesellschaft für Mathematik und
  Datenverarbeitung, St. Augustin, Federal Republic of Germany.

\bibitem[Reisig, 1991]{reisig1991algebraic}
Reisig, W. (1991).
\newblock Petri nets and algebraic specifications.
\newblock {\em Theoretical Computer Science}, 80:1--34.

\bibitem[Reisig, 2013]{reisig2013understanding}
Reisig, W. (2013).
\newblock {\em Understanding Petri Nets}.
\newblock Springer.

\bibitem[Reisig, 2019]{reisig2019associative}
Reisig, W. (2019).
\newblock Associative composition of components with double-sided interfaces.
\newblock {\em Acta Informatica}, 56(3):229--253.

\bibitem[Sandkuhl et~al., 2018]{sandkuhl2018modeling}
Sandkuhl, K., Fill, H.-G., Hoppenbrouwers, S., Krogstie, J., Leue, A., Matthes,
  F., Opdahl, A., Schwabe, G., Uludag, O., and Winter, R. (2018).
\newblock From expert discipline to common practice: A vision and research
  agenda for extending the reach of enterprise modelling, business and
  information systems engineering.
\newblock {\em Business and Information Systems Engineering}, 60(1):69--80.

\bibitem[Sanella and Tarlecki, 2012]{sanella20212algebraic}
Sanella, D. and Tarlecki, A. (2012).
\newblock {\em Foundations of Algebraic Specification and Formal Software
  Development}.
\newblock Springer.

\bibitem[Scheer, 1999]{scheer1999ARIS}
Scheer, A.-W. (1999).
\newblock {\em ARIS -- Business Process Frameworks}.
\newblock Springer, 3 edition.

\bibitem[Spivey, 1992]{spivey1992B}
Spivey, J.~M. (1992).
\newblock {\em The Z Notation: A reference manual}.
\newblock Prentice Hall, 2 edition.

\bibitem[Suppes, 1957]{suppes1957logic}
Suppes, P. (1957).
\newblock {\em Introduction to Logic}.
\newblock Van Nostrand Reinhold.

\bibitem[ter Hofstede and van~der Heide, 1992]{hofstede1992jungle}
ter Hofstede, A. H.~M. and van~der Heide, T.~P. (1992).
\newblock Formalization of techniques: chopping down the methodology jungle.
\newblock {\em Information and Software Technology}, 34(1):57--65.

\bibitem[Vernadate, 2020]{vernadat2020enterprise}
Vernadate, F. (2020).
\newblock Enterprise modelling: Research review and outlook.
\newblock {\em Computers in Industry}, 122:57--65.

\bibitem[Wand and Weber, 1990]{wand1990ontological}
Wand, Y. and Weber, R. (1990).
\newblock An ontological model of an information system.
\newblock {\em Transaction on Software Engineering}, 16(11):1282--1292.

\bibitem[Wand and Weber, 2002]{wand2002commentary}
Wand, Y. and Weber, R. (2002).
\newblock Research commentary: Information systems and conceptual modeling--a
  research agenda.
\newblock {\em Information Systems Research}, 13(4):363--376.

\bibitem[Winter, 2001]{winter2001engineering}
Winter, R. (2001).
\newblock Working for e-business -- the business engineering approach.
\newblock {\em International Journal of Business Studies}, 9(1):101--117.

\bibitem[Österle, 1995]{oesterle1995engineering}
Österle, H. (1995).
\newblock {\em Business in the Information Age -- Heading for New Processes}.
\newblock Springer.

\end{thebibliography}

\end{document}